\documentclass[11pt]{article}
\usepackage{amsmath}
\usepackage{graphicx}
\usepackage{lmodern}
\usepackage[colorlinks=true, linkcolor=blue, citecolor=blue, urlcolor=blue]{hyperref}
\usepackage{float} 
\usepackage{amssymb}
\usepackage{xcolor}
\usepackage{amsfonts}
\usepackage{mathtools}
\newcommand{\Lc}{${\Lambda_{\rm c}^+\;}$}
\newcommand{\D}{$\text{D}^0$}
\newcommand{\LcD}{${\Lambda_{\rm c}^+\rm /D^0\;}$}

\newcommand{\pt}{$p_{\rm T}\;$}
\newcommand{\ee}{$\rm e^-e^+$}
\newcommand{\ep}{$\rm e^-p\;$}

\newcommand{\ppb}{p--Pb\;}
\newcommand{\pp}{pp\;}
\newcommand{\lead}{Pb--Pb\;}
\newcommand{\dif}{differential\;}
\newcommand{\Lcpikp}{$\text{$\Lambda_{\text{c}}^+$} \rightarrow \text{p} \text{K}^- \text{$\pi$}^+$\;}
\newcommand{\Dkpi}{$\text{D}^0 \rightarrow \text{K}^-  \text{$\pi$}^+$\;}
\newcommand{\snn}{$\sqrt{s_{\text{NN}}} = 5.02 \; \text{TeV}\;$}
\usepackage[a4paper, margin=1in]{geometry} 

\usepackage[utf8]{inputenc}
\usepackage{microtype}

\title{Measurements of \LcD ratio as function of multiplicity at midrapidity at \snn}
\author{Oveis Sheibani$^{1,2}$ on behalf of ALICE collaboration}
\date{}

\begin{document}

\maketitle

\begin{center}
$^{1}$ University of Houston; osheibani@uh.edu \\
$^{2}$ ALICE Collaboration; oveis.sheibani@cern.ch\\
\vspace{10pt} 
{\small Hot Quarks 2022, Dao House, Estes Park, Colorado, USA, 10/11/2022 to 10/17/2022}
\end{center}

\begin{abstract}
In this contribution,  the measurement of prompt \LcD ratio as a function of multiplicity in \ppb collisions at mid-rapidity at \( \sqrt{s_{\rm_{NN}}} = 5.02 \) TeV is discussed. By performing this measurement as a function of multiplicity in \pp and \ppb collisions, we can evaluate the $p_{\rm T}$-differential baryon to meson enhancement and compare them to results in \ee and \ep collisions, where lower \LcD ratios at low and intermediate \pt have been observed, with the origin of this different behavior being still debated. In these measurements, we aim to compare the \ppb results to \pp collisions to investigate the possible effects of cold nuclear matter on charm-baryon production, and to \lead collisions for investigating the impact of quark--gluon plasma on charm quark hadronization.  \newline \newline Keywords: heavy-flavor, hadrons, baryons, mesons
\end{abstract}

\section{Introduction}
Measurements of heavy-flavor production in pp collisions are important to test perturbative quantum chromodynamics, since heavy quarks originate from hard scattering processes in the early stages of ultra-relativistic hadronic collisions. The measurements of the  $p_{\rm T}$-differential baryon-to-meson ratios in pp collisions at the LHC show a significant deviation from \ee and \ep collisions at low and intermediate \pt for several charm baryon species \cite{ALICE2022,Lisovyi2016}, suggesting that the hadronization mechanism of charm quarks in nucleon--nucleon collisions is modified with respect to \ee and \ep collisions. In \ppb collisions, the \LcD ratio shows a qualitatively similar trend with \pt as that measured in pp collisions, but the maximum is shifted toward higher \pt \cite{ALICE2022, ALICE2018}. In addition, the $p_{\rm T}$-differential ratio in pp collisions exhibits a significant dependence on the charged-particle multiplicity of the collision, with a $ 5.3\sigma $ deviation between the lowest and highest multiplicity bin \cite{ALICE2022}. In addition, the pp results at 13 TeV show a shift of the maximum of \LcD ratio to higher \pt in the highest multiplicity bin compared to the lowest multiplicity bin \cite{ALICE2022}, suggesting that charm hadron production in high-multiplicity collisions is influenced by collective effects and by a possible modification of the hadronization mechanism. It is important to study the multiplicity dependence of \LcD in \ppb collisions to further investigate such effects also in larger collision systems. This study, together with that in \pp collisions aims to understand the charm quark hadronization in small systems and evaluate possible similarities or differences, at large multiplicities, to the results in \lead collisions where a deconfined state of matter is expected to further affect the production of heavy flavor hadrons. 

\section{Analysis details}
The ALICE detector \cite{Aamodt2008} is used to measure the $\Lambda_c^+$ and $\rm D^0$ hadrons in \ppb collisions, reconstructed from their hadronic decay channels  \Lcpikp and \Dkpi, respectively, and selected exploiting their displaced decay topology and applying particle identification on their daughter tracks in a rapidity window of $-0.96<y<0.04$. The Inner Tracking System (ITS) is used for primary and secondary vertex reconstruction, the Time Projection Chamber (TPC) for track reconstruction and particle identification, and the Time Of Flight (TOF) detector for particle identification. 
The Bayesian combination of TPC and TOF is used to maximize the performance of particle identification. The V0M signals are used for the multiplicity estimation ${{\rm d}N_{\text{ch}}}/{{\rm d}\eta}$. The corrected yield of promptly produced (i.e. not originating from beauty-hadron decays) charm hadrons is calculated as: 
\begin{equation}
\frac{{\rm d} N^{\text{hadron}}_{\text{mult}}/{\rm d}p_{\rm T}}{N_{\text{mult}}^\text{ev}} \bigg|_{-0.96<y<0.04} = \frac{f_{\text{prompt}}(p_{\rm T}) \cdot \epsilon_{\text{trigger}} \cdot \frac{1}{2} N^{\text{hadron,raw}}_{\text{mult}}(p_{\rm T})|_{|y|<y_{\text{fid}}}}{N_{\text{mult}}^\text{ev}\Delta y  \Delta p_{\rm T}(Acc \times \epsilon)_{\text{prompt}}(p_{\rm T})(\text{BR})}
\end{equation}
where $N^\mathrm{hadron,raw}_\mathrm{mult}(p_T)$ is the value of the raw yield (sum of particles and antiparticles) extracted from the fit to the candidate invariant-mass distribution in the corresponding \pt and multiplicity interval in the fiducial rapidity range ($|y|<y_{\text{fid}}$). It is corrected for the beauty-hadron decay (feed-down) contribution ($f_{\text{prompt}}(p_{\rm T})$) and the trigger efficiency ($\epsilon_{\text{trigger}}$), divided by the acceptance-times-efficiency for prompt hadrons, $Acc \times \epsilon$, and divided by a factor of two to obtain the charge (particle and antiparticle) averaged yields. The yields were divided by the decay channel branching ratio (BR), the \pt interval width ($\Delta$\pt), and the rapidity coverage ($\Delta y$), and the number of events in each multiplicity bin ($N^{\text{ev}}_{\rm mult}$).

\section{Results}
\begin{figure}[H]
\includegraphics[width=16 cm]{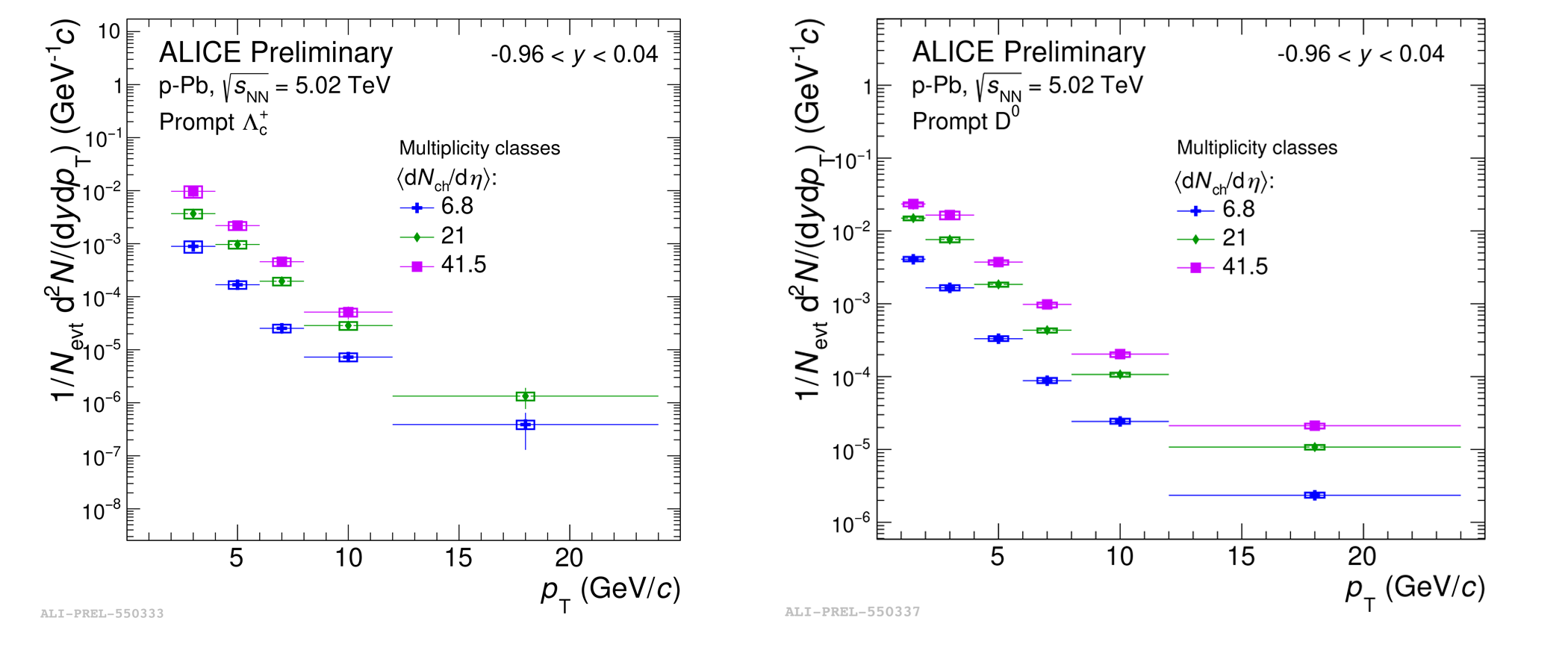}
\caption{$p_{\rm T}$-differential \Lc (left) and $\text{D}^0$ (right) corrected yields in three intervals of charged-particle multiplicity, in p--Pb collisions at \snn }
\label{fig1}
\end{figure}

\begin{figure}[H]
\includegraphics[width=16 cm]{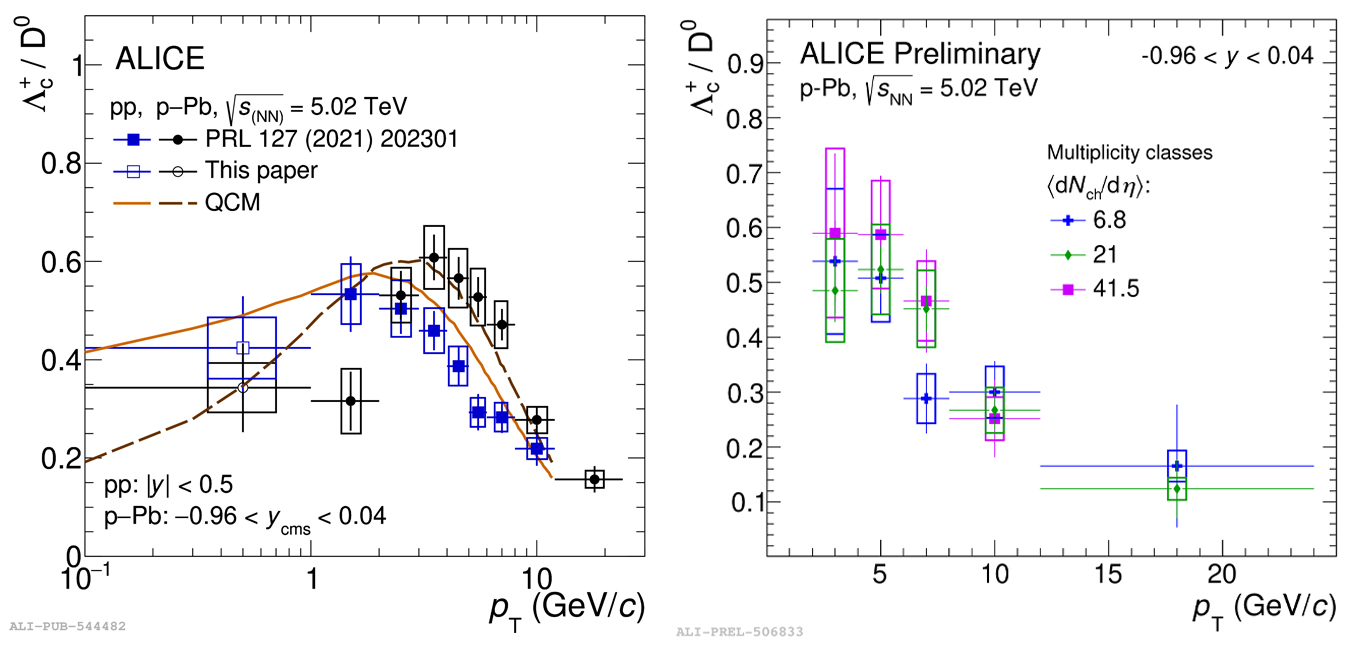}
\caption{$p_{\rm T}$-differential \LcD production yield ratios in p--Pb and pp collisions in minimum bias (left) \cite{down0} and for three intervals of charged-particle multiplicity (right) at \snn.}
\label{fig2}
\end{figure}   

\begin{figure}[H]
\includegraphics[width=16 cm]{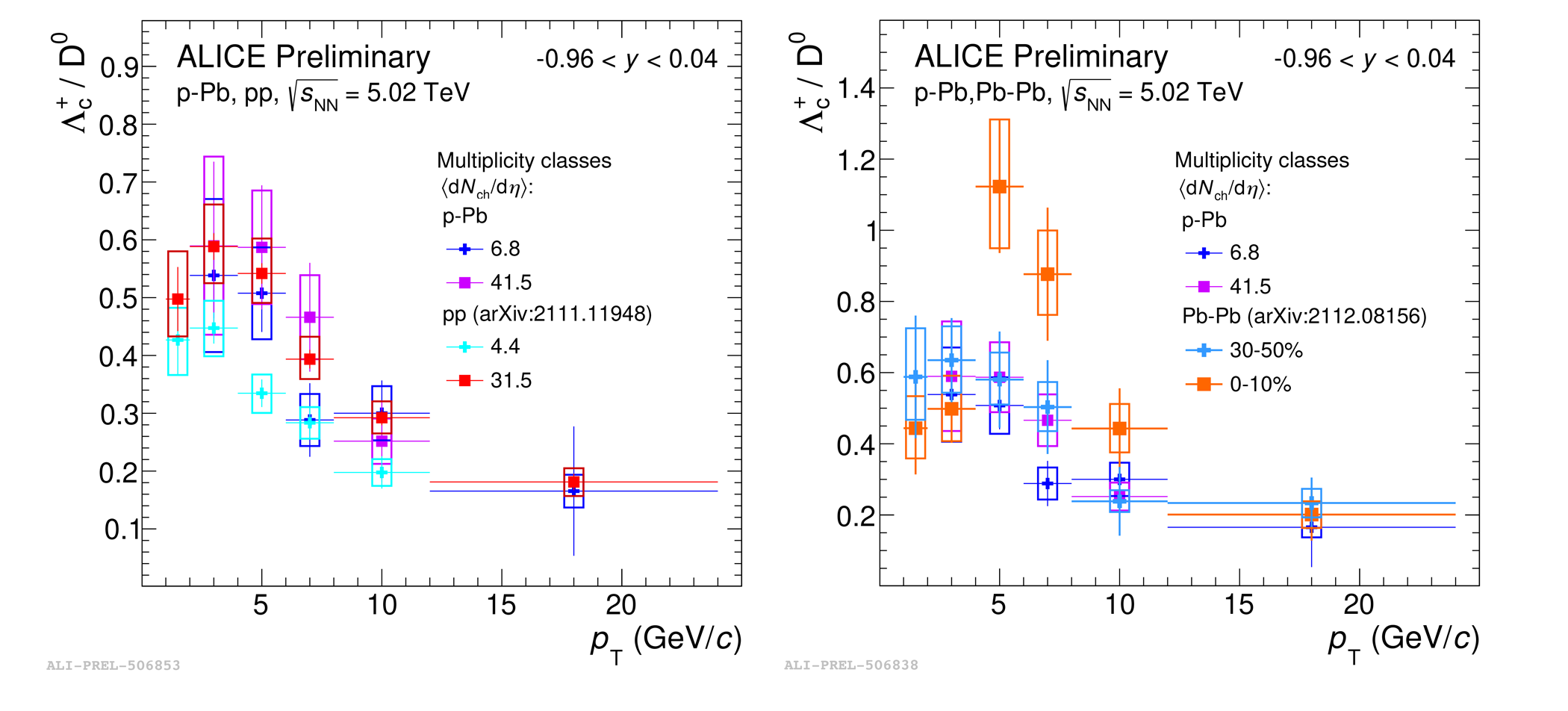}
\caption{$p_{\rm T}$-differential \LcD production yield ratios for three intervals of charged-particle multiplicity in p--Pb collisions at \snn compared with measurements in pp (left) and Pb--Pb collisions (right) at the same center-of-mass energy. }
\label{fig3}
\end{figure}   

\begin{figure}[H]
\includegraphics[width=16 cm]{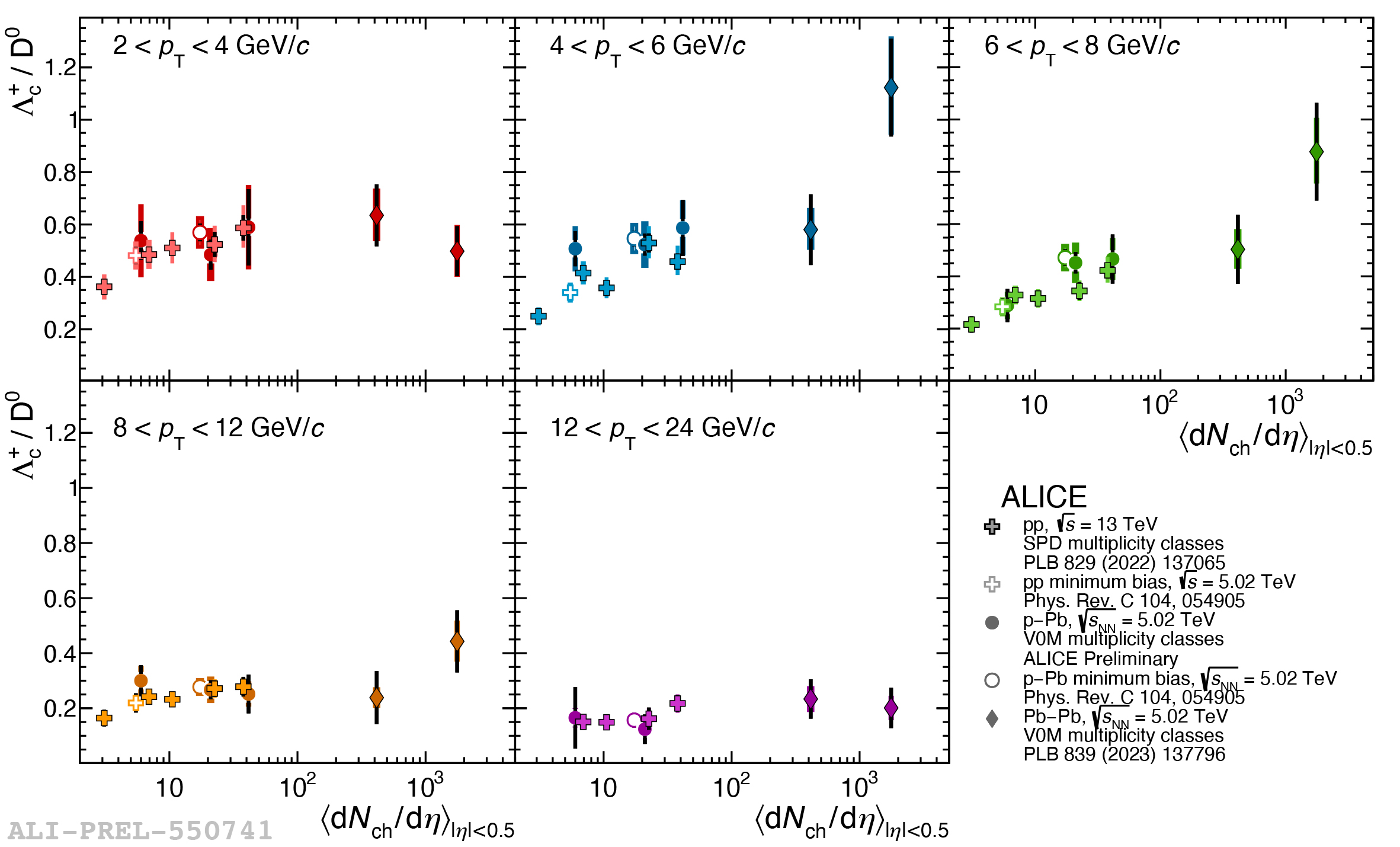}
\caption{$p_{\rm T}$-differential \LcD production yield ratios as a function of charged-particle multiplicity in pp, p--Pb, and Pb--Pb collisions at \snn
 }
\label{fig4}
\end{figure}

Figure ~\ref{fig1} shows the $p_{\rm T}$-differential corrected yield for \Lc baryons and \D mesons for \ppb collisions at \snn in different multiplicity intervals. By comparing the \LcD ratios in minimum bias \ppb and pp collisions, it can be observed that the maximum value of the \LcD ratio has shifted to higher \pt in \ppb compared to pp collisions (see Fig. \ref{fig2}, left) \cite{down0,Wilkinson2022}. 
In \ppb collisions, no significant multiplicity dependence is observed for the $p_{\rm T}$-differential ratios (see Fig. \ref{fig2} and Fig. \ref{fig4}). The highest and lowest multiplicity bins don't show discrepancies, differently from what observed in pp collisions. 
The results in semiperipheral (30-50\%) \lead collisions are compatible with those of the highest multiplicity interval in \ppb collisions. For the most central \lead collisions, the maximum of the \LcD ratio is shifted to higher \pt with respect to \ppb and pp, and the height of the peak is enhanced, exceeding unity at \( p_{\rm T} = 5 \) GeV/c (see Fig. \ref{fig3}, right). 

\begin{figure}[H]
\includegraphics[width=16 cm]{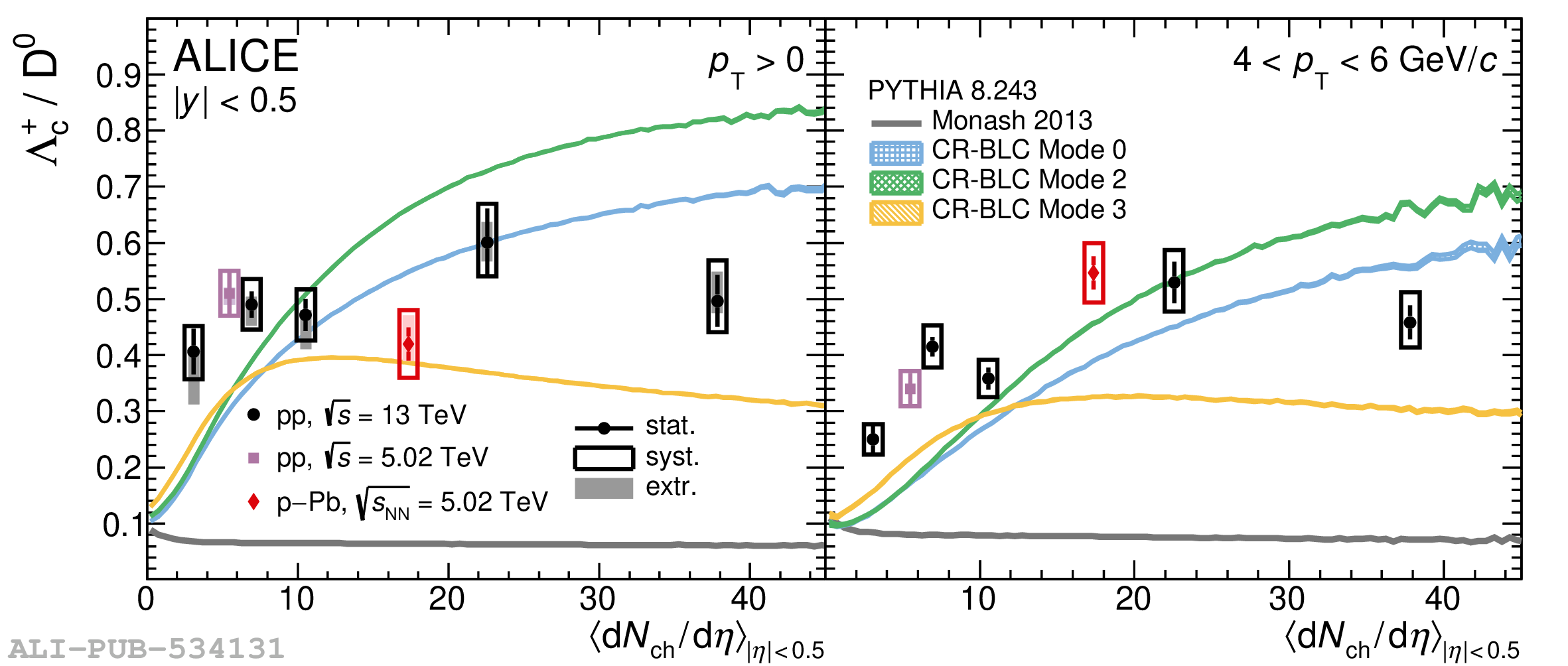}
\caption{$p_{\rm T}$-integrated \LcD0 yield ratios extrapolated for \pt $>$ 0 (left panel) and for the $4 < p_{\rm T} < 6\;$ GeV/c interval (right panel) }
\label{fig5}
\end{figure}   
\unskip

\section{Conclusion}
Models adopt multiple approaches to describe the observed modifications in pp collisions. For instance, PYTHIA8, by implementing a combination of color reconnections beyond leading color and multiparton interactions, can reproduce the \pt \dif results \cite{Weber2019, Skands2014}. 
The Catania model and the quark (re-)combination model, employ a combination of fragmentation and coalescence of partons within an expanding fireball, as described in \cite{Minissale2021, Song2018}, where radial flow plays a significant role. The Statistical Hadronization Model (SHM), with the inclusion of additional states from the relativistic quark model, can also reproduce the observed enhancement \cite{ALICE2022}. These observations strengthen the idea that other mechanisms in addition to fragmentation can be involved in charm (or bottom) hadronization, whose impact can vary as a function of the system size, here proxied by the event multiplicity. One can also speculate whether mechanisms such as radial flow might be responsible enhancing the baryon-to-meson ratio at certain \pt ranges. The $p_{\rm T}$-integrated \LcD ratios do not show any multiplicity dependence, suggesting the overall baryon to meson ratio could not be size dependent and the observed multiplicity evolution of the $p_{\rm T}$-differential \LcD ratios could be just due to a different redistribution of \pt between baryons and mesons for increasing multiplicities (see Fig. \ref{fig5}).



\end{document}